\documentclass[preprint,aps,pra,showpacs,eqsecnum]{revtex4}
\usepackage{amssymb}
\usepackage{amsmath}
\usepackage{graphicx}

\begin{document}

\title{The effect of measurements, randomly distributed in
time, on quantum systems. Stochastic quantum Zeno effect. }

\author{A. I. Shushin}
\affiliation{Institute of Chemical Physics, Russian Academy of
Sciences, 117977, GSP-1, Kosygin str. 4, Moscow, Russia}

\begin{abstract}
The manifestation of measurements, randomly distributed in time, on
the evolution of quantum systems are analyzed in detail. The set of
randomly distributed measurements (RDM) is modeled within the
renewal theory, in which the distribution is characterized by the
probability density function (PDF) $W(t)$ of times $t$ between
successive events (measurements). The evolution of the quantum
system affected by the RDM is shown to be described by the density
matrix satisfying the stochastic Liouville equation. This equation
is applied to the analysis of the RDM effect on the evolution of a
two level systems for different types of RDM statistics,
corresponding to different PDFs $W(t)$. Obtained general results are
illustrated as applied to the cases of the Poissonian [$W(t) \sim
e^{-w_r t}$] and anomalous [$W(t) \sim 1/t^{1+\alpha}, (\alpha \leq
1)$], RDM statistics. In particular, specific features of the
quantum and inverse Zeno effects, resulting from the RDM, are
thoroughly discussed.
\end{abstract}


\maketitle

\newpage

\section{Introduction}

The effect of measurements on the evolution of quantum systems has
recently been studied very actively both experimentally and
theoretically (for comprehensive review see, for example, refs.
\cite{Nak,Nam,Fac,Kos}). The interest to this problem was inspired
by the pioneering paper \cite{Mis} concerning the analysis of the
specific feature of the manifestation of measurements which was
called the Zeno effect, showing itself in the strong decrease of the
decay rate of the quantum state with the increase of a number of
measurements \cite{Mis}. Since then a lot of works, analyzing
different aspects of this effect, have been published.

Traditionally the quantum Zeno effect is considered assuming a set
of measurements to be distributed equidistantly in time (with
constant time between measurements). The equidistant distribution
results evidently in simplification of the mathematical treatment of
the problem and experimental observation of the measurement effect
\cite{Nak,Nam,Fac,Kos}.

Naturally, the majority of manifestations of the measurement effect
(including the Zeno effect), found for equidistant distribution of
measurements, are expected to occur in the case of irregular
distribution as well. However, some additional analysis is certainly
needed.

In this work we will discuss the measurement effect in the
interesting special case of the irregular distribution, the case of
measurements randomly distributed in time, called hereafter randomly
distributed measurements (RDM). The random process of measurements
is modeled within the renewal approach (RA) which treats the
sequence of measurements as a stochastic set of renewals
\cite{Wei,Bou}. In the RA the distributions of time intervals $t$
between successive renewals are assumed to be stochastically
independent and are described by the probability density function
(PDF) $W(t)$ (often denoted as $\psi (t)$ \cite{Wei}).

The evolution of the quantum system, affected by the RDM of this
type, is shown to be described by the stochastic Liuvilles equation
(SLE) for the density matrix of the system \cite{Shu1,Shu2}. The SLE
allows one to analyze the effect of the RDM for different types PDFs
$W(\tau)$ in a fairly simple analytical form.

In our work general expressions for density matrix of the system,
affected by the RDM are derived. With the use of these expressions
some important specific features the RDM effect are analyzed for
different types of $W(t)$ behavior. In particular, the quantum Zeno
effect, i.e. the decrease of the decay rate of the state with
decreasing the average time between measurements $\bar t$, is
predicted only in the case of rapidly decreasing PDFs $W(t)$ for
which the average time between measurements $\bar t =
\int_0^{\infty}\! d\tau\, \tau W(\tau)$ is finite. As to heavy
tailed PDFs, which cannot be described by a finite $\bar t$ ($\bar t
\to \infty$), for these PDFs the quantum Zeno effect is shown to be
absent. Such a drastic difference of the RDM effect in these two
cases appears to be conveniently treated in terms of the Zeno-effect
dependence on the characteristic rate $w_r$ of the Laplace transform
$\widetilde{W} (\epsilon)$ as a function of the Laplace variable
$\epsilon$. The fact is that, unlike $\bar t$, the rate $w_r$ can be
introduced independently of the mathematical form of $W(t)$
decrease, as it will be shown in our work, though for rapidly
decreasing $W(t)$ these two parameters are closely related: $\bar t
\sim 1/w_r$.

To illustrate the general results we will discuss the case of the
Poissonian distribution with $W(t)= e^{-w_r t}$, as an example of
the RDM with rapidly decreasing PDF, and the case of anomalous heavy
tailed distribution $W (t) \sim 1/t^{1+\alpha}$ with $\alpha \leq
1$.

In particular, within the Poissonian model some characteristic
properties of the RDM effect are studied which are typical for case
of rapidly decreasing PDF $W(t)$. These properties manifest
themselves in specific features of the time dependence of the
probability $p(t)$ to survive in the measured state. It is shown,
for instance, that in the limit of small characteristic time $t_r =
w_r^{-1}$ between measurements $p(t)$ is the exponentially
decreasing function with the rate non-monotonically depending on
$t_r$: the decreasing dependence at very small $t_r$, corresponding
to the quantum Zeno effect, is changed by the increasing one,
associated with the inverse Zeno effect \cite{Nam,Fac,Kos,Fac1}, as
$t_r$ is increased. The time $t_{r_m}$ of the change is determined
by the parameters of system.

In both Poissonian and anomalous RDM models the manifestation of
relaxation and, for example, the appearance and specific features of
the quantum and inverse Zeno effects in the presence of relaxation
are also analyzed be means the proposed approach.

In conclusion, possible realistic examples of processes, in which
the predicted effects can be observed, are thoroughly discussed. In
particular, some realizations of both Poissonian and anomalous RDM
in the processes with the participation of Brownian particles are
considered.

\section{Formulation of the problem}

In this work we consider the effect of the RDM on the evolution of
dynamic quantum systems, i.e. the systems, in which relaxation is
absent. The statistical properties of the RDM are treated within the
renewal theory \cite{Wei,Bou}, well known in statistical physics.

In order to analyze the RDM effect we, first, should clarify some
details of this effect for the finite number, say, $n$ of
measurements.

\subsection{Measurement affected quantum dynamics}

The dynamic quantum system is characterized by the wave function
$|\psi (t) \rangle$ (the vector in the Hilbert space) whose time
evolution is governed by the Schr\"odinger equation ($\hbar = 1$)
\begin{equation}\label{mes0}
|\dot \psi \rangle = -iH |\psi \rangle, \;\;\mbox{with} \;\; H = H_0
+ H_i.
\end{equation}
In this equation $H$ is the Hamiltonian of the system represented as
a sum of the free and interaction parts, $H_0$ and $H_i$,
respectively. The initial condition for eq. (\ref{mes0}) $|\psi (t =
0) \rangle = |\psi_i \rangle$ depends on the process considered (see
below).

The time evolution of the wave function is usually described by
expansion in some (complete) basis of functions $|\psi_j \rangle$ in
the Hilbert space. In our further analysis, for the sake of
convenience, we will assume $|\psi_j \rangle$ to be basis of
eigenfunctions of the free Hamiltonian $H_0$.

In accordance with the conventional von Neumann rule
\cite{Nam,Fac,Kos,Von}, measurements, which show that the system is
in some state suggested to be the eigenstate of $H_0$ with the wave
function $|\psi_0 \rangle$ ($H_0 |\psi_0 \rangle = \omega_0 |\psi_0
\rangle$), enable one to obtain the probability $p (t)$ to find the
system in the state $|\psi_0 \rangle$: $p (t) = |\langle \psi_0
|\psi (t) \rangle |^2$.

In what follows we will analyze the probability $p(t) $ of survival
in the state $|\psi_0 \rangle$ after a set of measurements.
Following refs. \cite{Nak,Fac} the subtle problem of the evaluation
of the multiple measurement effect will be treated assuming that "if
every time the measurement has a positive outcome and the system is
found in the initial state, the wave function "collapses" and the
evolution starts anew from $|\psi_0 \rangle$". In this case the
problem is to calculate the probability $p(t)$ to find the system,
initially created in $|\psi_0 \rangle$ (i.e. with $|\psi_i \rangle =
|\psi_0 \rangle$), in the same state $|\psi_0 \rangle$ after the
measurements.

It is worth noting that, in fact, the effect of measurements reduces
to population and phase relaxation in the system under study (see
below). In such a case, in general, the evolution of the system
should be described with the density matrix $\rho (t)$ satisfying
the Schr\"edinger like equation but in the Liouville space
\cite{Von}. In this space the matrix $\rho (t)$ is represented as a
vector in the basis, consisting of bilinear combinations of wave
functions $|\psi_j \rangle$
\begin{equation}\label{mes1}
|jj' \rangle = |\psi_j \rangle \langle \psi_{j'}|:
\end{equation}
\begin{equation} \label{mse1a}
\rho (t) \equiv |\rho (t)\rangle = \sum\nolimits_{\nu = jj'}
\rho_{\nu} (t)|\nu\rangle.
\end{equation}

Note that in the Liouville space for any vector $|\nu \rangle, \,
(\nu = jj'),$  the corresponding conjugated one $\langle \nu|$ is
defined by the relation $\langle \nu|\nu' \rangle =
\delta_{\nu\nu'}$. Noteworthy is also that vectors $|jj' \rangle$
and $|j'j \rangle$ are considered as independent and $\langle
jj'|j'j \rangle = 0$.

For the dynamic system, whose wave function $|\psi(t) \rangle$
satisfies eq. (\ref{mes0}), the density matrix is represented as
$\rho (t) = |\psi(t) \rangle \langle \psi (t) |$. The above
mentioned Shr\"odinger equation (in the Liouvulle space) for this
matrix is similar to eq. (\ref{mes0}) but with the Hamiltonian
operator $H$ replaced by the superoperator $\hat H$:
\begin{equation}\label{mes2}
\dot \rho =  -i\hat H \rho, \;\;\mbox{where} \;\; \hat H \rho =
[H,\rho]\equiv H\rho -\rho H.
\end{equation}
According to the definition (\ref{mes1}) the matrix elements of the
superoperator $\hat H$ are given by $\langle kk'| \hat H |jj'
\rangle = \langle \psi_k|H|\psi_j \rangle \delta_{j'k'}  - \langle
\psi_{j'}|H |\psi_{k'} \rangle \delta_{jk} $. It is important to
note that for the Hermitian Hamiltonian $H$ the superoperator $\hat
H$ is also Hermitian (but in the Liouville space).

The effect of a measurement on the evolution of the quantum system
is conveniently determined in terms of the projection superoperator
$\hat P_m = |00 \rangle \langle 00 |$, where $|00 \rangle = |\psi_0
\rangle \langle \psi_0 |$. In particular,  after $n$ successive
measurements at times $\{t_j\} (j = 0,\dots,n)$ ordered as $t \geq
t_n \geq t_{n-1} \geq \dots \geq t_1 \geq t_0 = 0$, the density
matrix $\rho (t)$ can be obtained with formula
\begin{equation}\label{mes3}
\rho (t)  = \hat U_{{\bf t}_n}\!(t) \rho_0, \, (n = 0, 1,\dots),
\end{equation}
in which ${\bf t}_n = (t_n, \dots, t_0)$ is the vector of
measurement times, $\hat U_{{\bf t}_0}\!(t) = \hat U_0(t-t_0)$ and
\begin{equation}\label{mes3a}
\hat U_{{\bf t}_{n\geq 1}}\!(t) = \hat U_0(t-t_n)\prod_{j=1}^{n}
[\hat P_m \hat U_0(t_{j}-t_{j-1})]
\end{equation}
with $\hat U_0 (t) = e^{-i\hat Ht}$ being the evolution
superoperator (operator in the Liouville space), and $\rho_0 =
|\psi_0 \rangle \langle \psi_0 | \equiv |00 \rangle$.

With the density matrix the observable under study, i.e. the
probability of survival in the state $|\psi_0 \rangle$ after $n$
measurements, is expressed as
\begin{equation}\label{mes4}
p_n(t)
= \langle \psi_0 |\rho(t)|\psi_0  \rangle = 
{\rm Tr}[\hat P_m \hat U_{{\bf t}_{n}}\!(t)],
\end{equation}
where the trace is evaluated over the states $|jj' \rangle$ in the
Liouville space.

The probability $p_n (t)$ is known to depend not only on the
parameters of the Hamiltonian $H$ but also on the number $n$ of
measurements. The dependence is thoroughly analyzed in a large
number of publications  assuming of fixed time interval $\tau_m =
t/n$ between measurements (see, for example, reviews
\cite{Nam,Fac,Kos,Von}). Below we will extend the analysis assuming
the multiple measurements to be a stochastic processes (in general
non-Markovian), which will be modeled within the RA \cite{Wei,Bou}.

\subsection{Renewal approach}

\subsubsection{General formulas}

In the RA the times $t_n$ of events (measurements) are considered to
be randomly distributed in time with intervals $\tau_{n} = t_n -
t_{n-1}$ between successive events described as independent random
variables with the monotonically decreasing PDFs $W(\tau_n)$, the
same for all intervals. Note that $W(t)$ is defined only for
positive $t$, with $W(t < 0) = 0$. To completely characterize the
statistics of events one also needs the probability $P(t)$ that the
interval between the successive events is greater than $t$: $P(t) =
\int_{t}^{\infty}\!dt' W(t')$, which is, naturally, normalized by
the relation $P (0) = 1$.

In what follows we will mainly operate with the Laplace transforms
denoted as
\begin{equation}\label{ren0}
{\widetilde Z}(\epsilon) = \int_0^{\infty} \! dt \, Z(t)e^{-\epsilon
t}
\end{equation}
for any function $Z(t)$. In particular, noteworthy is the relation
$\hat {\widetilde P} (\epsilon)= [1 - \hat {\widetilde W}
(\epsilon)]/\epsilon$ and suitable representations
\begin{equation}\label{ren1}
{\widetilde {W}} (\epsilon) = [1 + \Phi (\epsilon)]^{-1} \;\:
\mbox{and} \;\: {\widetilde P} (\epsilon) = [\epsilon + \epsilon/
\Phi (\epsilon)]^{-1}.
\end{equation}
in terms of the  auxiliary function $\Phi (\epsilon)$

The analytical form of $\Phi (\epsilon)$ is completely determined by
that of $W(t)$. In what follows, to specify the characteristic scale
of $\Phi (\epsilon)$-dependence we will introduce the characteristic
rate $w_r$ whose meaning will become clear from some particular
examples of functions $\Phi (\epsilon)$ considered below. It is
important that this rate can be introduced both for rapidly and
anomalously slowly decreasing PDFs $W(t)$ (see below).

The effects analyzed in this work are essentially controlled by the
small $\epsilon$ behavior of $\Phi (\epsilon)$. In general, we can
only say that $\Phi (\epsilon) \stackrel{\epsilon \to 0} \to 0$.
Some additional information on $\Phi (\epsilon)$ behavior at
$\epsilon \to 0$ can be obtained by the analysis of the long time
dependence of $W(t)$:

a) If at $t \to \infty$ the PDF $W(t)$ decreases so rapidly that
average time (first moment) $\bar t = \int_{0}^{\infty} \! dt \, t
W(t) \sim w_r^{-1}$ is finite, at small $\epsilon$ the function
$\Phi (\epsilon)$ can be represented as \cite{Wei}
\begin{equation}\label{ren1a}
\Phi (\epsilon) \stackrel{\epsilon \to 0}{=} \epsilon \bar t +
o(\epsilon/w_r).
\end{equation}

b) If, however, $W(t)$ is a heavy tailed function: $W(t\to\infty)
\to 1/t^{1+\alpha}$ with $\alpha < 1$, and, therefore, $\bar t$ is
infinite (does not exist), then $\Phi (\epsilon) \stackrel{\epsilon
\to 0}{\approx} (\epsilon/w_r)^{\alpha}$ \cite{Wei,Bou}.

Within the RA, due to independence of renewals, the PDF $W_{{\bf
t}_n} (t)$ of $n$ events ($n \geq 0$) at times $\{t_j\} (j = 0,
1,\dots,n)$, satisfying the relation $t_n \geq t_{n-1} \geq \dots
\geq t_1 \geq t_0 = 0$ and combined into the vector ${\bf t}_n =
(t_n,\dots,t_1)$, is given by
\begin{equation} \label{ren2}
W_{{\bf t}_{n\geq 1}} = \prod_{j=1}^{n} W(t_j - t_{j-1}).
\end{equation}

These functions completely describe the stochastic renewal process.
In particular, with the use of $W_{{\bf t}_n}$ the probability
$\pi_n (t)$ to observe $n$ events in the time interval $(0,t)$ is
expresses by \cite{Wei}:
\begin{equation} \label{ren2a}
\pi_{n} (t) =  \int_0^{t} \! dt' \, P(t-t')W_n (t') dt', \;\; (n
\geq 0).
\end{equation}
In this formula $W_{n}(t)$ is the PDF of $n$ events, which for $n
\geq 2$ is equal to $W_{{\bf t}_{n}}$ integrated over all $t_{j\leq
n-1}$ (from $0$ to $t_{j+1}$, respectively) except for $t_n = t$,
and for $n = 0, 1$ are defined by $W_{1}(t) = W(t),$ and $W_{0}(t) =
\delta (t)$. The functions $\pi_n (t)$ can, evidently, be
represented as
\begin{equation} \label{ren3}
W_{n} (t) =  \frac{1}{2\pi i}\int_{-i\infty}^{i\infty} \! d\epsilon
\,{\widetilde {W}}^{n} (\epsilon) e^{\epsilon t} d\epsilon.
\end{equation}
Note that according to eq. (\ref{ren3}) $W_{0} (t)=\delta (t)$ and
therefore $\pi_0 (t) = P(t)$. It addition, it is worth noting, that
the probabilities $\pi_n (t)$ satisfy the normalization condition
$\sum_{n=0}^{\infty}\pi_n = 1$.

\subsubsection{Examples of renewal processes}

In our further analysis special attention will be paid to some
particular renewal processes corresponding to different distribution
of times of renewals, i.e. different functional form of the PDF
$W(t)$.

\paragraph{Poissonian distribution.}

The most simple is the Poissonian model, corresponding to
\cite{Wei,Bou}
\begin{equation} \label{ren6}
P (t) = e^{-w_r t}, \;\; W (t) =  w_r e^{-w_r t},
\end{equation}
so that ${\widetilde {W}} (\epsilon) = [1 + (\epsilon/w_r)]^{-1}$
and therefore
\begin{equation} \label{ren7}
\Phi (\epsilon) = \epsilon/w_r.
\end{equation}

\paragraph{Equidistant distribution.}

The model of equidistant distribution, in which
\begin{equation} \label{ren4}
P (t) = \theta (t_r - t),\;\; W (t) = \delta (t - t_r) ,
\end{equation}
with $t_r = 1/w_r$, describes the set of events with the constant
interval $t_r$ between successive ones \cite{Shu3}. This is just the
model considered in almost all publications concerning the Zeno
effect. In this model
\begin{equation} \label{ren5}
\Phi (\epsilon) = e^{t_r \epsilon} - 1.
\end{equation}

\paragraph{Anomalous distribution.}

The anomalous model implies the L\'evy-type distribution
\cite{West1} of times between events with the heavy tailed behavior
of $W (t) \sim 1/t^{1+\alpha},\, (\alpha \leq 1)$. One can find a
number of such type of models which predict the same results for
long time features of processes. In our analysis we use the simple
one, for which \cite{Wei,Bou,Shu1}
\begin{equation} \label{ren8}
P (t) = E_{\alpha} (-(w_r t)^{\alpha}), \; W (t) =  -\dot P (t),\;
(\alpha \leq 1),
\end{equation}
where $E_{\alpha} (-x^{\alpha}) = (2\pi i)^{-1}
\int_{-i\infty}^{i\infty}\!dz \, e^{xz} (z + z^{1-\alpha})^{-1}$ is
the Mittag-Leffler function \cite{West1}. This model corresponds to
\begin{equation} \label{ren9}
\Phi (\epsilon) = (\epsilon/w_r)^{\alpha}.
\end{equation}
It is easily seen that the expression (\ref{ren8}) predicts the
heavy tailed behavior the PDF $W (t)$.

\section{Stochastic Liouville equation}

The above consideration shows that the RDM effect on the evolution
of the quantum system under study is expressed in terms of the the
superoperator $\hat U (t)$ [eq. (\ref{mes3a})] averaged over the
stochastic process of measurements. In what follows this averaged
superoperator will be denoted as $\hat {\overline{U}}(t)$.

The problem of finding $\hat {\overline{U}}(t)$, which is a
functional of type of (\ref{mes3a}) averaged over the renewal
process, has already been discussed in literature \cite{Shu1,Shu2}.
It is shown to reduce to solving some equation for this operator,
which is called the SLE.

For some types of renewal process the averaging is essentially
simplified. The most well known example is the Poissonian process,
simplification for which results from the Markovian nature of this
process. In the RDM analysis the Poissonian process of quantum
transitions (jumps) is just a realization of the Markovian process
of "migrations" (or jumps) over states $|jj'\rangle$  of the quantum
system under study (in the Liovulle space). The treatment of this
Poissonian "migration" process within the widely accepted continuum
time random walk approach \cite{Mont1,Mont2} shows that the PDF
$\rho_M$ for the "migrating" system satisfies the Markovian equation
$\dot \rho_M = -w_r (1 - \hat P_m)\rho_M$. The effect of such
"migrations" on quantum evolution is described by the SLE for the
evolution suproperator  $\dot {\hat {\overline{U}}}(t)= - iH \hat
{\overline{U}}(t) - w_r (1 - \hat P_m)\hat {\overline{U}}(t)$
\cite{Kub} (see Sec. IV.C).

Fortunately, the SLE can be obtained for any type of renewal
process. In the most general form the SLE is rigorously derived with
the Markovian representation of the RA \cite{Shu1,Shu2}. In this
work, however, we will restrict ourselves to the simplest variant of
the RA, for which the SLE can be obtained fairly easily. Below we
will outline some details of the derivation.

The derivation is based on the fact that in the absence of
measurements ($n = 0$) the evolution operator $\hat U_{{\bf t}_0}(t)
= \hat U_{0}(t)$, while for any number $n \geq 1$ of measurements
the operator $\hat U_{{\bf t}_n}(t)$ [eq.(\ref{mes3a})] and the PDF
of measurements $W_{{\bf t}_n}$ (\ref{ren2}) are represented as
products of terms depending on differences of times $t_j - t_{j-1}$.
In such a case we get the following formulas for corresponding
average evolution operators $\hat {\overline{U}}_{n}(t)$: $\hat
{\overline{U}}_{0}(t) = P(t)\hat U_{0}(t)$, for $n = 0$, and
\begin{equation} \label{sle1}
\hat {\overline{U}}_{n}(t) = \int_0^{t} \! d{\bf t}_n \,
P(t-t_n)W_{{\bf t}_n} \hat U_{{\bf t}_n}(t),\;\;\mbox{for} \: n\geq
1,
\end{equation}
where $d{\bf t}_n = \prod_{j=1}^{n} dt_j$. The convolution-like form
of formulas for $\hat {\overline{U}}_{n} (t)$ results in a simple
representation for the Laplace transforms of these functions:
\begin{equation} \label{sle2}
\hat {\widetilde{\overline{U}}}_{n}(\epsilon) = \widetilde{P}(\hat
\Omega_{\epsilon})\big[ \hat P_m\widetilde{W} (\hat
\Omega_{\epsilon})\big]^n \;\;\mbox{with}\;\; \hat \Omega_{\epsilon}
= \epsilon + i \hat H .
\end{equation}

Summing up the contributions for different numbers of measurements
we thus get the evolution operator $\hat {\overline{U}}(t)$ averaged
over the renewal process \cite{Shu1,Shu2}
\begin{eqnarray}
\hat {\widetilde{\overline{U}}} (\epsilon) = \sum_{n=0}^{\infty}
\hat {\widetilde{\overline{U}}}_{n}(\epsilon) \!&=& \widetilde{P}(
\hat \Omega_{\epsilon}) \big[1-P_m\widetilde{W}
(\hat \Omega_{\epsilon})\big]^{-1} \label{sle3a}\\
&=&\hat \Omega_{\epsilon}^{-1}\Phi (\hat \Omega_{\epsilon})\big[\Phi
(\hat \Omega_{\epsilon}) + \hat Q_m)\big]^{-1}.\quad \label{sle3b}
\end{eqnarray}
where $\hat Q_m = \hat 1 - \hat P_m = \sum_{jj'\neq 00} |jj'
\rangle\langle jj'|$ is the superoperator of projection onto the
subspace $\{|jj'\rangle \}, \, (jj' \neq 00)$.

The expression (\ref{sle3b}) can be treated as a solution of the
non-Markovian SLE for $\hat {\overline{U}}(t)$
\begin{equation} \label{sle5}
\dot {\hat{{\overline{U}}}} = -i\hat H {\hat {\overline{U}}} - \hat
Q_m \! \int_0^{t}\!\! d\tau \, M(\tau) e^{-i \hat H\tau} \hat
{\overline{U}}(t-\tau)
\end{equation}
in which
\begin{equation} \label{sle6}
M(t) = \frac{1}{2\pi i}\, \frac{d}{dt}\int_{-i\infty}^{i\infty}\!\!
d\epsilon \, e^{\epsilon t} \Phi^{-1} (\epsilon).
\end{equation}
is the memory function whose analytical properties are essentially
determined by those of the PDF $W(t)$.

The SLE (\ref{sle5}) is not quite convenient for applications. In
our further analysis we will mainly use the expression (\ref{sle3b})
for the Laplace transform of the evolution operator.

In accordance with the general formula (\ref{mes4}) the (average)
probability $p(t)$ of survival in the initial state ($|\psi_0
\rangle$) is completely determined by the average evolution operator
$\hat{{\overline{U}}} (t)$. In terms of the Laplace transforms the
corresponding expression is written as
\begin{equation} \label{sle6a}
\widetilde{p} (\epsilon) = {\rm Tr}[\hat P_m \hat
{\widetilde{\overline{U}}}(\epsilon)] \equiv {\rm Tr}[\hat P_m\hat
{\widetilde{\overline{U}}}(\epsilon)\hat P_m],
\end{equation}
where, similar to formula (\ref{mes4}), trace is evaluated over the
states $|jj' \rangle$ in the Liouville space.

In the conclusion of this general analysis we will show that with
the use of above-obtained formulas $\widetilde{p} (\epsilon)$ can be
expressed in terms of the probability $p_0 (t) = \langle 00| \hat
U_0 (t)|00\rangle = \langle 00| e^{-i\hat H t}|00\rangle$ of
survival in the initial state $|\psi_0\rangle$ in the absence of
measurements. According to eq. (\ref{sle6a}) the Laplace transform
$\widetilde{p} (\epsilon)$ is determined by the trace of the
supermatrix
\begin{equation} \label{sle6b}
\hat {\widetilde{\overline{U}}}_{P}(\epsilon) \equiv \hat P_m \hat
{\widetilde{\overline{U}}}(\epsilon)\hat P_m =
\sum\nolimits_{n=0}^{\infty}\hat
{\widetilde{\overline{U}}}_{P_n}(\epsilon),
\end{equation}
where $ \hat {\widetilde{\overline{U}}}_{P_n}(\epsilon) = \hat
P_m\hat {\widetilde{\overline{U}}}_{n}(\epsilon) \hat P_m $. Each
term $\hat {\widetilde{\overline{U}}}_{P_n}(\epsilon)$ in the sum
can be represented by formula
\begin{equation} \label{sle6b0}
\hat {\widetilde{\overline{U}}}_{P_n}(\epsilon) = \hat
{\!\widetilde{P}}_P(\epsilon) {\hat
{\,{\widetilde{W}}_{\!p_0^{}}^n}} (\epsilon).
\end{equation}
In this expression we have introduced the supermatices
\begin{equation} \label{sle6b1}
\hat {\widetilde{X}}_{P}(\epsilon) = \hat P_m\widetilde{X}(\hat
\Omega_{\epsilon})\hat P_m \;\; \mbox{for}\; X = P,\: W,
\end{equation}
which are directly related to the probability $p_0 (t)$, as is clear
from the relations $\hat P_m\widetilde{X}(\hat
\Omega_{\epsilon})\hat P_m = \int_0^{\infty} dt \, e^{-\epsilon t}
\hat P_m e^{-i\hat H t}\hat P_m X(t)$ and $\hat P_m e^{-i\hat H
t}\hat P_m = \hat P_m p_0 (t)$. Finally we obtain
\begin{equation} \label{sle6b2}
\hat {\widetilde{\overline{U}}}_{P_n}(\epsilon) = \hat P_m
{\widetilde{P}}_{p_0}(\epsilon) {\widetilde{W}}_{\!p_0}^n
(\epsilon),
\end{equation}
where
\begin{equation} \label{sle6b3}
{\widetilde{X}}_{p_0}(\epsilon) = \int_0^{\infty}\!dt \,
e^{-\epsilon t} X(t) p_0 (t)\;\; \mbox{for} \;\; X = P, \,W.
\end{equation}
Substitution of formula (\ref{sle6b2}) into eqs. (\ref{sle6b}) and
(\ref{sle6a}) yields
\begin{equation} \label{sle6b4}
\widetilde{p} (\epsilon) =
{\widetilde{P}}_{p_0}(\epsilon)[1-{\widetilde{W}}_{\!p_0}
(\epsilon)]^{-1}.
\end{equation}

In principle, both formulas (\ref{sle6a}) [with eq. (\ref{sle3b})]
and (\ref{sle6b4}) are quite suitable for the analysis of the RDM
effect on $p (t)$. In what follows, however, we will mainly apply
the formulation based on the first formula [(\ref{sle6a})], though,
eq. (\ref{sle6b4}) will also be used to clarify some particular
properties of the effect.

\section{Zeno effect on two level systems}

In this section we will analyze the effect of the RDM on some model
quantum system to illustrate the specific features of manifestation
and treatment of the Zeno effect in this kind of measurements.

The simplest (though quite realistic) system, which enables one to
significantly simplify mathematical problems, is the two level
system. It is very convenient for detailed description of all
important features of the Zeno effect \cite{Nak,Nam,Fac,Kos}).

\subsection{Hamiltonian of the two level system}

In our analysis we will use the Hamiltonian in the form [see eq.
(\ref{mes0})]
\begin{equation} \label{sle7}
H_0 = \varepsilon (|1\rangle \langle 1|-|2\rangle \langle 2|), \;\;
H_i = v (|1\rangle \langle 2| + |2\rangle \langle 1|).
\end{equation}
in which $\varepsilon$ and $v$ are a positive real parameters. The
measured state is assumed to be the state $|1\rangle$, i.e. $|\psi_0
\rangle = |1\rangle$.

In the Liouville space the Hamiltonian is represented as a $4 \times
4$ matrix
\begin{eqnarray}
\hat H_0 &=& 2\varepsilon (|12\rangle \langle 12|-|21\rangle \langle
21|),
\label{sle8a}\\
\hat H_i &=& v [(|11\rangle - |22\rangle)(\langle 21| - \langle 12|)
\nonumber\\
&& + (|21\rangle - |12\rangle)(\langle 11| - \langle 22|)]
\label{sle8b},
\end{eqnarray}
in the basis $|jj'\rangle = |j\rangle \langle j'|$ [defined in eq.
(\ref{mes1})]. In this basis the superoperator $\hat P_m$,
describing the effect of a measurement, is written as
\begin{equation} \label{sle10}
\hat P_m = |11\rangle \langle 11| \;\;\mbox{and}\;\;\hat Q_m =
\sum\nolimits_{jj' \neq 11} |jj' \rangle \langle jj'| .
\end{equation}

The Laplace transform of the function under study, the survival
probability $\widetilde{p} (\epsilon) $, can conveniently be
evaluated with the expression (\ref{sle6a}), which, as applied to
the two level system considered, is written as
\begin{equation} \label{sle11}
\widetilde{p} (\epsilon) = {\rm Tr}[\hat P_m \hat
{\widetilde{\overline{U}}}(\epsilon)] = \langle 11|\hat
{\widetilde{\overline{U}}}(\epsilon) |11\rangle.
\end{equation}

With the above formulas at hand one can analyze the specific
features of the Zeno effect for any type of the RDM.

\subsection{General results}

Here we will obtain some general results, valid for quantum systems
with arbitrary (but finite) number of levels, to clarify the
manifestation of the analytical properties of the decreasing
function $W (t)$ in the specific features of the Zeno effect.

The most important property of function $W (t)$ is the rate of
decrease at long times which is determined by the behavior of $\Phi
(\epsilon)$ in the limit $\epsilon \to 0$ (Sec. IIB.1), i.e at
$\epsilon \ll w_r$, where $w_r$ is the above defined rate,
characterizing the time $t_r = 1/w_r$ of the onset of the asymptotic
long time behavior of $W(t)$. Just in this limit, or more accurately
in the limit of large $w_r$, when $\xi = \|\hat
\Omega_{\epsilon}\|/w_r \ll 1$, one can demonstrate some important
properties of the RDM effect. In principle, essential conclusions
can be made without any particular assumptions on $\Phi
(\epsilon)$-behavior except for the relation $\Phi (\epsilon)
\stackrel{\epsilon \to 0} \to 0$. However, in the analysis it is
suitable to keep in mind the approximation $\Phi (\epsilon \to 0)
\approx (\epsilon/w_r)^{\alpha}$, with $\alpha \leq 1$, which is of
particular interest for our further discussion.

The above mentioned important general conclusions concern the
properties of $\widetilde{p} (\epsilon) $ at $\xi  \ll 1$. In this
limit we get from eqs. (\ref{sle3b}) and (\ref{sle11})
\begin{eqnarray}
\widetilde{p} (\epsilon) &\approx& \widetilde{p}_{\infty} (\epsilon)
= {\rm Tr} [\hat P_m \hat \Omega_{\epsilon}^{-1}\Phi (\hat
\Omega_{\epsilon})]/\langle 11| \Phi (\hat
\Omega_{\epsilon}) |11\rangle\nonumber\quad\\
&=& \langle 11|\hat \Omega_{\epsilon}^{-1}\Phi (\hat
\Omega_{\epsilon})|11\rangle/\langle 11| \Phi (\hat
\Omega_{\epsilon}) |11\rangle.\qquad \label{sle11b}
\end{eqnarray}
This general expression is valid for systems with any number of
levels, though it is presented in terms of the considered two level
model (\ref{sle8a}),(\ref{sle8b}).

Formula (\ref{sle11b}) can be derived from eqs. (\ref{sle3b}) and
(\ref{sle11}) by taking into account some characteristic properties
of the matrix $\hat L = \Phi (\hat \Omega_{\epsilon}) + \hat Q_m$.
For brevity, in our further study we will use the notation $\Phi
(\hat \Omega_{\epsilon}) = \hat \Phi$. To clarify the derivation we
will, first, analyze the specific features of two parts of $\hat L$:
$\hat L_P = \ \hat P_m \hat L \hat P_m = \hat P_m \hat \Phi \hat
P_m$ and $\hat L_Q = \hat Q_m \hat L \hat Q_m = \hat Q_m \hat \Phi
\hat Q_m + \hat Q_m$, operating in subspaces $\{|11\rangle\}$ and
$\{|jj'\neq 11\rangle\}$, respectively. The fact is that in the
considered limit $\xi = \|\hat \Omega_{\epsilon}\|/w_r \ll 1$,
corresponding to $\| \hat \Phi \| \ll 1$, the eigenvalue $\lambda_P
= \langle 11|\hat \Phi |11\rangle \ll 1$ of the matrix $\hat L_P$ is
much smaller than $\hat L_Q$-eigenvalues $\lambda_{Q_{\nu}} \sim 1$,
whose magnitudes are mainly determined by $\hat Q_m$. These
estimations yield for the characteristic splitting of eigenstates
$\delta \lambda = \lambda_{Q_{\nu}} - \lambda_P$ the value $\delta
\lambda \sim 1$.

The obtained splitting appears to be much larger than the $\hat
L$-induced interaction $ \hat L_{QP} = \hat P_m \hat L \hat Q_m +
\hat Q_m \hat L \hat P_m = \hat P_m \hat \Phi \hat Q_m + \hat Q_m
\hat \Phi \hat P_m $ between the states of $\{|11\rangle\}$ and
$\{|jj'\neq 11\rangle\}$ subspaces: $\| \hat L_{QP} \| \sim \| \hat
\Phi\| \ll \delta \lambda$. This means that in the leading order in
$\| \hat \Phi\|/\delta\lambda \sim \| \hat \Phi\| \ll 1$ the
eigenstates and eigenvalues of the matrix $\hat L$ coincide with
those of $\hat L_P + \hat L_Q$. Of special importance is the
coincidence of the lowest eigenstate $|l\rangle$ of $\hat L$ with
$|11\rangle$: $|l\rangle \sim |11\rangle + \sum_{jj'\neq
11}\zeta_{jj'}|jj'\rangle$ with $\zeta_{jj'} \sim \| \hat \Phi\| \ll
1$, which results in formula $\langle l| \hat L |l\rangle = \langle
11|\hat L_P|11\rangle [1 + O(\| \hat \Phi\|)]$ with $\langle 11|\hat
L_P |11\rangle = \langle 11|\hat \Phi |11\rangle$. This formula
implies that with the accuracy $\sim \| \hat \Phi\| \ll 1$ the
Green's function $\hat L^{-1}$ is mainly determined by the
contribution of the eigenstate $|l\rangle \approx |11\rangle$ :
\begin{equation} \label{sle11bb}
\hat L^{-1} = \hat P_m \langle 11|\hat \Phi |11\rangle^{-1} [1 +
O(\| \hat \Phi\|)].
\end{equation}
Other eigenstates (which with high accuracy coincide with those of
$\hat L_Q$) make a contribution much smaller than that of
$|l\rangle$, since the corresponding eigenvalues $\lambda_{Q_{\nu}}
\sim 1$ are much larger than $\lambda_P = \langle 11|\hat \Phi
|11\rangle \ll 1$.

Substitution of the expression (\ref{sle11bb}) into eqs.
(\ref{sle3b}) and (\ref{sle11}) yields formula (\ref{sle11b}). Note
that in the same way  it can also be derived by means of eq.
(\ref{sle6b4}).

This formula allows one to analyze the limiting behavior of the RDM
effect in the limit $w_r \to \infty$, i.e. for very small
characteristic time between measurements.

In particular, we can discuss the case of rapidly decreasing $W (t)$
for which $\Phi (\epsilon) \stackrel{\epsilon \to 0}{\approx}
\epsilon \bar t$ [see eq. (\ref{ren1a})]. Taking into account the
relation (\ref{ren1a}) one arrives at the estimation $\widetilde{p}
(\epsilon) = \epsilon^{-1}[1 + O(\|\hat \Omega \|/w_r)]$ which means
that
\begin{equation} \label{sle11c}
p(t) \stackrel{w_r \to \infty} \longrightarrow 1.
\end{equation}

The limiting  relation (\ref{sle11c}) demonstrates the localization
of the system in the measured state in the limit $w_r \to \infty$,
which is associated with the quantum Zeno effect \cite{Nam,Fac,Kos})
expressed in terms of the RA.

The dependence of $p (t)$ on the average time $\bar t \sim w_r^{-1}$
between measurements is, in general, not universal. It is also
determined by other parameters of the system, for example, the
parameters of the Hamiltonian. In such a case, to characterize the
quantum Zeno effect as a function of $w_r^{-1}$, one can apply the
average time $t_Z (w_r) = \int_0^{\infty}\!dt\, p(t) = \widetilde{p}
(\epsilon=0)$:
\begin{eqnarray} \label{sle11d}
t_Z &=&  \langle 11|\hat \Omega_{\epsilon}^{-1}\Phi (\hat
\Omega_{\epsilon})\big[\Phi (\hat \Omega_{\epsilon}) + \hat
Q_m)\big]^{-1} |11\rangle\big|_{\epsilon \to 0}\qquad \nonumber\\
&=&{\widetilde{P}}_{p_1}(\epsilon)[1-{\widetilde{W}}_{\!p_1}
(\epsilon)]^{-1}\big|_{\epsilon \to 0}.
\end{eqnarray}
where $\Omega_{\epsilon} = \epsilon + i\hat H$. The second of eqs
(\ref{sle11d}) is written with the use the expression (\ref{sle6b4})
for $\widetilde{p} (\epsilon=0)$ in which, however, the functions
${\widetilde{P}}_{p_0}(\epsilon)$ and ${\widetilde{W}}_{p_0}
(\epsilon)$ are replaced with ${\widetilde{P}}_{p_1}(\epsilon)$ and
${\widetilde{W}}_{p_1}(\epsilon)$, respectively, defined by
${\widetilde{X}}_{p_1}(\epsilon) = \int_0^{\infty}\!dt \,
e^{-\epsilon t} X(t) p_1 (t)\;\; \mbox{for} \;\; X = P, \,W$. This
replacement is made due to the change of notation for the measured
state ($|1\rangle$ instead of  $|0\rangle$), according to which, to
avoid possible confusions, $p_0 (t)$ should be replaced by $p_1 (t)
= \langle 11| e^{-\hat H t} |11 \rangle$.

In general, $t_Z (w_r)$ can be calculated only numerically. Some
conclusions on the specific features of this parameter, however, can
be made by the analysis of limiting behavior  of $t_Z (w_r)$ at $w_r
\to \infty$ and $w_r \to 0$.

It is seen from the above definition of ${\widetilde{P}}_{p_1}
(\epsilon)$ and ${\widetilde{W}}_{\!p_1} (\epsilon)$ that the value
of $t_Z$ is finite in the case of finite $\bar t = \int_0^{\infty}
dt\, tW(t) = \int_0^{\infty} dt\, P(t)$. In general, $t_Z$ can be
calculated only numerically. However, some conclusions on the
specific features of this parameter can be made by the analysis of
limiting behavior of $t_Z (w_r)$ at $w_r \to \infty$ and $w_r \to
0$.

1) For  $w_r \to \infty$ we get the relation $t_Z (w_r \to \infty)
\to \infty$ which follows from the definition of this parameter and
the relation $\widetilde{p} (\epsilon) = \epsilon^{-1} $ valid in
this limit.

2) In the opposite limit $w_r \to 0$ the parameter $t_Z (w_r)$ also
grows to infinity: $t_Z (w_r \to 0) \sim 1/w_r \to \infty$. This
dependence can be obtained by analyzing $w_r$-dependence of
${\widetilde{P}}_{p_1} (\epsilon = 0)$ and ${\widetilde{W}}_{p_1}
(\epsilon = 0)$ at small $w_r$. The fact is that the probability
$p_1 (t) = \langle 11| e^{-\hat H t} |11 \rangle$, which determines
the values of these two functions, can, in general, represented as:
$p_1 (t) = \bar p_1 + \delta p_1 (t)$, where $\bar p_1$ is
independent of time and $\delta p_1 (t)$ is the oscillating part
represented as a sum of harmonically oscillating functions.  The
most important for our analysis is $\bar p_1$, which can be found by
expansion of the evolution operator $e^{-i Ht}$ in the basis of
eigenfunctions $|\varphi_j\rangle $ of the Hamiltonian $H$: $\bar
p_1 = \sum_j |\langle \psi_1|\varphi_j\rangle|^4 < 1$. With this
representation for $p_1(t)$ one can obtain the estimations
${\widetilde{P}}_{p_1}(0) \approx \bar p_1 \int_0^{\infty}\!dt \,
P(t) = \bar p_1 \bar t = \bar p_1 /w_r$ and
${\widetilde{W}}_{p_1}(0) \approx \bar p_1 \int_0^{\infty}\!dt \,
W(t) = \bar p_1 < 1$, in which the contribution of the oscillating
part $\delta p_1 (t)$, negligibly small in the limit $w_r \to 0$, is
ignored. Substitution of these relations into eq. (\ref{sle11d})
leads to the above-mentioned limiting dependence $t_Z (w_r \to 0)
\approx \bar p_1 (1-\bar p_1)^{-1}w_r^{-1}$.

The above analysis shows that the behavior of $t_Z (w_r)$ is
non-monotonic with the minimum of this function at some $w_r$ whose
value is determined by the parameters of system. The validity of
such a conclusion will be demonstrated below as applied to the case
of the Poissonian RDM distribution, as an example.

\subsection{Poissonian distribution of measurements}

In the case of the Poissonian RDM statistics, when $\Phi (\hat
\Omega_{\epsilon}) = \hat \Omega_{\epsilon}/w_r $, equation
(\ref{sle3b}) for the Laplace transform ${\widetilde{\overline{U}}}
(\epsilon)$ is essentially simplified reducing to the SLE of
Schr\"odinger type \cite{Kub}
\begin{equation} \label{poi1}
(\hat \Omega_{\epsilon} + w_r\hat Q_m)\hat
{\widetilde{\overline{U}}} (\epsilon) \equiv (\epsilon + i\hat H +
w_r\hat Q_m) \hat {\widetilde{\overline{U}}} (\epsilon) = 1.
\end{equation}
which is briefly discussed in Sec. III.

For the two level system this equation is the system of four linear
equations which can be solved analytically. We are not going to
present the cumbersome expressions for matrix elements of $\hat
{\widetilde{\overline{U}}} (\epsilon)$ but restrict ourselves to
obtaining only one element corresponding to the observable under
study, the survival probability $p (t)$ [see eq. (\ref{mes4})]. For
the initial condition $|\psi_0\rangle = |1\rangle$ the Laplace
transform $\widetilde{p} (\epsilon)$ is given by
\begin{eqnarray} \label{poi2}
\widetilde{p} (\epsilon) &=& \langle 11| \hat
{\widetilde{\overline{U}}} (\epsilon) |11\rangle \nonumber \\
&=& \frac{\epsilon  + w_r+ \bar w(\epsilon)}{\epsilon^2 + \epsilon [
w_r + 2 \bar w(\epsilon)] + w_r \bar w(\epsilon)}\,,
\end{eqnarray}
where
\begin{equation} \label{poi3}
\bar w (\epsilon) = 2(\epsilon + w_r)v^2/[(\epsilon + w_r)^2 +
4\varepsilon^2].
\end{equation}

Of special interest is, naturally, the limiting variant of general
formula (\ref{poi2}) corresponding to large average rate  $w_r$ of
repetition of measurements. It is seen from this expression, that in
the limit $w_r \gg v$ (and for $\epsilon < w_r$) $\widetilde{p}
(\epsilon) \approx (\epsilon + \bar w_0)^{-1}$, i.e. for $t >
w_r^{-1}$,
\begin{equation} \label{poi3a}
p(t) \approx e^{-\bar w_0 t}\;\;\mbox{with}\;\; \bar w_0 = \bar w
(0) = \frac{2 w_rv^2}{w_r^2 + 4\varepsilon^2}.
\end{equation}
Note that in the considered limit $w_r \gg v$ the rate $\bar w_0
\sim w_r (v/w_r)^2 \ll w_r$.

Equation (\ref{poi3}) shows that in the limit of large $w_r$ the
decay of the survival probability turns out to be exponential with
the rate $\bar w_0$, in principle, non-monotonically depending on
$w_r$. The non-monotonic behavior, however, can can correctly be
described by eq. (\ref{poi3a}) only in the case $\varepsilon \gg v$,
when within the region of validity of this formula ($w_r \gg v$)
there exists the subregion of $w_r$ values, $\varepsilon \gg w_r \gg
v$, in which $\bar w_0 (w_r)$ is the increasing function of $w_r$:
$\bar w_0 (w_r) \sim w_r$. The non-monotonic dependence $\bar w_0
(w_r)$ can be treated as the acceleration of $p(t)$-decay by
measurements at relatively low measurement rates $w_r \ll
\varepsilon$ (associated with the inverse Zeno effect) followed by
the deceleration at large $w_r \gg \varepsilon$: $\bar w_0 (w_r)
\sim 1/w_r$ (corresponding to the quantum Zeno effect).

In general, the specific features of the RDM effect can be
demonstrated with the parameter $t_Z (w_r )= \widetilde{p} (0)$. The
expression (\ref{poi2}) yields simple formula for this function:
\begin{equation} \label{poi4}
t_Z = w_r^{-1} + \bar w_0^{-1} = v^{-1}[1 +
\mbox{$\frac{1}{2}$}(\bar w_{r}^2 + 4\bar \varepsilon^2)]/\bar w_r
\end{equation}
with $\bar \varepsilon = \varepsilon/v$ and $\bar w_{r} = w_r/v$. In
agreement with general qualitative conclusions (Sec. IVB), $t_Z
(w_r)$ non-monotonically depends on $w_r$ with $t_Z \stackrel{w_r
\to 0}{\approx} 1/w_r \to \infty$, $t_Z \stackrel{w_r \to
\infty}{\approx} w_r/(2v^2) \to \infty$, and the minimum at
\begin{equation} \label{poi4a}
\bar w_{r_m} = (2 + 4 \bar\varepsilon^2)^{1/2}.
\end{equation}
For  $\varepsilon \gg v$ in the limit $w_r \gg v$ the non-monotonic
behavior of $t_Z (w_r ) \approx \bar w_0^{-1}$ with $\bar w_{r_m} =
2\bar\varepsilon$ indicates the occurrence of two regimes of the
Zeno effect, quantum and inverse, mentioned above. It is seen that
in the case $\varepsilon \gg v$ the coordinate $\bar w_{r_m} $($=
2\bar\varepsilon$) predicted by eq. (\ref{poi4a}) coincides with
that of the maximum of the rate $\bar w_{0} (w_r)$. The results of
this analysis demonstrate that the parameter $t_Z (w_r)$ is
certainly useful for studying qualitative specific features of the
RDM effect.

Characteristic properties of the behavior of the probability $p(t)
\equiv p(\tau|\tau_r)$ (at fixed $\tau = tv$) as a function of the
dimensionless average time $\tau_r = \bar w_r^{-1} = v/w_r$ between
measurements are shown in Fig. 1.

At very small $\tau_r$ the probability $p(\tau|\tau_r)$
monotonically approaches $1$ as $\tau_r \to 0$, in agreement with
predictions of the quantum Zeno effect. The above analysis with the
SLE (\ref{poi1}) clearly reveals the mechanism of the slowing down
of $1 \rightarrow 2$ transitions by the RDM. According to this
equation the RDM effect on the evolution of the state $|1\rangle$ is
actually equivalent to the effect of the decay of the state
$|2\rangle$ with the rate $w_r$, which is accompanied by {\it
dephasing} (decay of the density matrix elements $\langle 1 |\rho
|2\rangle$ and $\langle 2 |\rho |1\rangle$) with the same rate
$w_r$. Just fast dephasing in the limit $w_r \to \infty$ leads to
the strong reduction of the $1 \rightarrow 2$ transition rate [$\sim
v (v/w_r)$], associated with the quantum Zeno effect.

At intermediate values of $\tau_r$ the function $p(\tau|\tau_r)$
non-monotonically depends on $\tau_r$ with the minimum at some
$\tau_{r_{Pm}} = v/w_{r_{Pm}}$. The position of the minimum is
reasonably accurately estimated by means of $\bar w_{r_m}$
(\ref{poi4a}) for all values of $\bar \varepsilon = \varepsilon/v$
used in numerical calculation: $\tau_{r_{Pm}} = v\tau_{r_{Pm}}
\approx \bar w_{r_m}^{-1}$. The accuracy of this estimation is
especially good for largest considered value $\bar \varepsilon =
2.5$, as expected from eq. (\ref{poi3a}) and above analysis.

The analysis of the expression (\ref{poi3a}) makes it possible to
understand the reason of the appearance of this minimum. It results
from interplay between quantum oscillations in the system and RDM
induced dephasing. Just this interplay manifests itself in the
factor $w_r^2 + 4\varepsilon^2$ in the denominator in formula
(\ref{poi3a}) for $\bar w_0$, which is responsible for the
non-monotonic behavior of $p(\tau_r)$.

In the region of validity of eq. (\ref{poi3a}), i.e. for $\tau_r =
v/w_r \ll 1$, the kinetics of $p(\tau)$ decay is exponential. This
means that the non-monotonic behavior of the dependence $p(\tau_r)$
at a fixed $\tau = vt$ stems from that of the rate $\bar w_0 (\bar
w_r)$. Such a dependence of the rate is traditionally treated as a
transition from the quantum Zeno effect [in the region of decreasing
behavior of $\bar w_0 (w_r)$] to the inverse Zeno effect [when $w_0
(w_r)$ is the increasing function] \cite{Fac}. In general, however,
for non-exponential $p (\tau)$-dependence, it is not quite correct
to associated the non-monotonic behavior of $p(\tau_r)$ with any
specific phenomena because of a large number of possible
peculiarities of this behavior $p (\tau)$-kinetics in general.

\subsection{Anomalous distribution of measurements}

Of special interest is the case of the anomalous RDM, in which the
PDF $W(t)$ anomalously slowly decreases at large times $t$: $W(t)
\stackrel{t\to\infty}{\sim} 1/t^{1+\alpha}$ with $\alpha < 1$. It is
easily seen that this PDF cannot be characterized by the average
time $\bar t$ between measurements in its usual definition though,
of course, this functions still has the characteristic decay time
(see below). In our further consideration we will apply the simplest
Mittag-Leffler model for the anomalous PDF $W(t)$, defined in eqs.
(\ref{ren8}) and (\ref{ren9}). The existence and qualitative
definition of the characteristic time $t_r = w_r^{-1}$ in this model
is clear from eq. (\ref{ren9}).

For brevity in our analysis of the anomalous case we will restrict
ourselves to the most interesting limit of high characteristic
inverse time between measurements $w_r$, corresponding to
$\|\hat\Omega\|/w_r \ll 1$. In this limit $\widetilde{p}(\epsilon)
\approx \widetilde{p}_{\infty} (\epsilon)$ can be evaluated with
formula (\ref{sle11b}), which in the applied Mittag-Leffler model is
represented as \cite{Shu4}
\begin{eqnarray}
\widetilde{p}_{\infty} (\epsilon)  &=& \langle 11|\hat
\Omega_{\epsilon}^{\alpha -1}|11\rangle/\langle 11| \hat
\Omega_{\epsilon}^{\alpha} |11\rangle \label{anom1a}\nonumber\\
&=& \frac{(2\bar \varepsilon^2+1) \,\epsilon^{\alpha-1}+
\bar{\Omega}^{\alpha-1}_{\epsilon}} {(2\bar\varepsilon^2+1)
\,\epsilon^{\alpha}+\bar{\Omega}^{\alpha}_{\epsilon}},\qquad
\label{anom1b}
\end{eqnarray}
where $\bar\varepsilon = \varepsilon/v$ and
\begin{equation} \label{anom2}
\bar \Omega^{\beta}_{\epsilon} =
\mbox{$\frac{1}{2}$}\big[\big(\epsilon + 2iv\sqrt{\bar\varepsilon^2
+ 1}\big)^{\beta} + \big(\epsilon - 2iv\sqrt{\bar\varepsilon^2 +
1}\big)^{\beta}\big].
\end{equation}

It is easily seen that for the Poissonian RDM ($\alpha = 1$)
$\widetilde{p}_{\infty} (\epsilon) = 1/\epsilon$, so that
$p_{\infty} (t) = 1$ as predicted in presence of the quantum Zeno
effect.

As for the anomalous RDM, the existence of the non-trivial limit
$\widetilde{p}_{\infty} (\epsilon)$ itself indicates the violation
of the charactristic Zeno-effect behavior of $\widetilde{p}_{\infty}
(\epsilon)$ in this case. In the limit $\|\hat\Omega\|/w_r \ll 1$
the function $\widetilde{p}_{\infty} (\epsilon)$ [and thus $p (t) =
p_{\infty} (t)$] appears to be independent of $w_r$ and determined
by the parameters of the Hamiltonian only.

At short times $t \ll \varepsilon^{-1}, v^{-1}$ formula
(\ref{anom1b}) predicts no transitions, i.e. $\widetilde{p}_{\infty}
(\epsilon) \approx 1/\epsilon$ and $p(t) \approx 1$. In the limit of
long times $t \gg \varepsilon^{-1}, v^{-1}$, however, one gets
\begin{equation} \label{anom3}
p_{\infty}(t)\sim
A_{\alpha}(t) t^{-\alpha},
\end{equation}
where $A_{\alpha} (t)$ is the oscillating function of time: $
A_{\alpha} (t) = a_{\alpha} + c_{\alpha}\cos (2\bar E t) +
s_{\alpha}\sin (2\bar E t)$, in which $\bar E = v
\sqrt{\bar\varepsilon^2 + 1}$ and $a_{\alpha}$, $c_{\alpha}$, and
$s_{\alpha}$ are the constants depending on $\alpha$ and
$\bar\varepsilon$. This asymptotic expression can be derived by
taking into account that the most slowly decreasing (and additive)
contributions to the integral of the inverse Laplace transformation
$p_{\infty}(t) = (2\pi i)^{-1}\int_{-i\infty}^{i\infty} d\epsilon\,
\widetilde{p}_{\infty} (\epsilon) e^{\epsilon t}$ come from
singularities of $\widetilde{p}_{\infty} (\epsilon)$ (\ref{anom1b})
located at the imaginary axis of the complex plane of $\epsilon$.
The singularities (brunching points) are determined by the terms
$\epsilon^{\alpha-1}$ and $\bar{\Omega}^{\alpha-1}_{\epsilon}$ in
the numerator in eq. (\ref{anom1b}).  In the long time limit $t \gg
1/\sqrt{\varepsilon^{2} + v^2}$ these singularities contribute
independently and the evaluation of contributions reduces to the
calculation of integrals of type of $(2\pi
i)^{-1}\int_{-i\infty}^{i\infty} d\epsilon\, \epsilon^{\alpha-1}
e^{\epsilon t} \sim t^{-\alpha}$, which leads to eq. (\ref{anom3}).

It is important to note that the transition from the anomalous case
$\alpha < 1$ to the Poissonian one $\alpha = 1$ is fairly
non-trivial. In order to clarify the details of this transition we
will consider the case of $\alpha$ close to $1$, when
$\delta_{\alpha} = 1-\alpha \ll 1$. In this limit the onset of the
inverse power type kinetics (\ref{anom3}) displaces to very long
times and the major part of the kinetics $p(t)$ reduces to the
exponential one, which can be obtained by expanding
$\widetilde{p}_{\infty} (\epsilon)$ in powers of $\delta_{\alpha}$.
For example, taking into account that for $\delta_{\alpha} \ll 1$
$x^{\alpha - 1} \approx x (1 - \delta_{\alpha} \ln x)$, one can
represent the numerator as $(2\bar \varepsilon^2+1)
\,\epsilon^{\alpha-1}+ \bar{\Omega}^{\alpha-1}_{\epsilon} = 2(\bar
\varepsilon^2+1) [1 + O(\delta_{\alpha})]$. As for the denominator,
similar expansion results in $(2\bar\varepsilon^2+1)
\,\epsilon^{\alpha}+\bar{\Omega}^{\alpha}_{\epsilon} = 2(\bar
\varepsilon^2+1)(\epsilon + w_z) [1 + O(\delta_{\alpha})]$. Finally
one gets $\widetilde{p}_{\infty} (\epsilon) \approx (\epsilon +
w_z)^{-1}$ and therefore
\begin{equation} \label{anom4}
p_{\infty}(t)\approx e^{-w_z t}\;\;\mbox{with}\;\; w_z =
\mbox{$\frac {1}{2}$} \pi (1-\alpha) v\sqrt{\bar\varepsilon^2 + 1}.
\end{equation}
It is worth noting that the characteristic behavior (\ref{anom4}) is
determined by that of $\widetilde{p}_{\infty} (\epsilon)$ at small
$|\epsilon | \sim \delta_{\alpha}$. In this region the terms of
higher order in $\delta_{\alpha} = 1-\alpha$, whose summarized
contribution is denoted as $O(\delta_{\alpha})$, result in the
correction of the kinetics $\sim \delta_{\alpha}^2 \ln
(1/\delta_{\alpha})$ which can be neglected in the limit
$\delta_{\alpha} \ll 1$. These terms, non-analytical in $\epsilon$,
are responsible for the inverse time behavior of $p_{\infty}(t)$ at
long times.

Figure 2 shows the dependence $p_{\infty}(t)$ for some particular
sets of parameters of the model.

\section{Effect of relaxation}

\subsection{General remarks}

So far we have considered the RDM effect on dynamical systems only,
though it is known that the relaxation can strongly modify the
manifestation of this effect \cite{Nam,Fac,Kos}). In particular, in
addition to the quantum and inverse Zeno regimes of the effect,
analyzed above, there appears another one, in which no strong
influence of measurements on the evolution kinetics is observed in
the limit of short average time $\bar t \sim w_r^{-1} \to 0$. This
regime can also be considered as a manifestation the inverse Zeno
effect \cite{Fac,Kos} though this is a matter of definition.

The influence of relaxation on the Zeno effect has already been
analyzed within the model of equidistantly distributed measurements
\cite{Nam,Fac,Kos}). Here we will discuss this influence in the case
of the RDM.

In general, the analysis of the relaxation in quantum systems is a
difficult problem. In this work we will restrict ourselves to the
simple Markovian case, in which the kinetic equation for the density
matrix can be written as
\begin{equation} \label{rel1}
\dot \rho  = -\hat {\cal L} \rho, \;\;\mbox{where} \;\; \hat {\cal
L} = i\hat H + \hat R.
\end{equation}
Here $\hat R$ is the relaxation superoperator which can be evaluated
in the short correlation time (or the Bloch-Redfield) approximation
\cite{Abr}. Some example of the relaxation model, leading to the
particular expression for the superoperator $\hat R$ will be
considered below.

It is easily seen that in the presence of relaxation all general
formulas are similar to those, obtained above for dynamic systems.
The only difference is in the definition of $\Omega_{\epsilon}$. In
the presence of relaxation
\begin{equation} \label{rel2}
\hat \Omega_{\epsilon} = \epsilon  + \hat {\cal L} = \epsilon + \hat
R + i\hat H.
\end{equation}

To apply the results obtained in Sec. III with the redefined
$\Omega_{\epsilon}$  we need to specify the relaxation
superoperator, i.e. the relaxation mechanism.

\subsection{Simple relaxation model}

In order to illustrate the main features of the manifestation of
relaxation in the RDM effect it is sufficient to analyze the
simplest relaxation models. In our analysis we will discuss the
variant of the model, widely accepted in the magnetic resonance
theory \cite{Abr}, in which
\begin{eqnarray} \label{rel3}
\hat R &=& w_d (| 11\rangle  - | 22 \rangle) (\langle 11 | -
 \langle 22 |)\qquad \nonumber\\
&& + w_p( | 12 \rangle \langle 12 |+ | 21\rangle \langle 21 |) .
\end{eqnarray}
This operator describes population relaxation with the rate $w_d$
and dephasing with the rate $w_p$. The rates satisfy the relation
$w_p \geq \frac{1}{2} w_d$, which ensures positivity of the density
matrix $\rho (t)$ during the evolution, described by eq.
(\ref{rel1}) \cite{Hab}.

The model (\ref{rel3}) allows one to analyze fairly easily the
important properties of the relaxation effect for any type of the
PDF $W(t)$.

\subsubsection{Poissonian distribution of measurements}

As we have already mentioned above, the case of the Poissonian RDM
can be considered as a simple example of renewal processes with
rapidly decreasing PDF $W(t)$. The substitution of $\hat
\Omega_{\epsilon}$ (\ref{rel2}) into eq. (\ref{poi1}) leads to the
expression (\ref{poi2}) for $\widetilde{p} (\epsilon)$, but with
$\bar w (\epsilon)$ replaced by $\bar w_d (\epsilon) = \bar w
(\epsilon) + w_d$ and with $w_r$ changed by $w_r + w_p$ in the
function $\bar w (\epsilon)$ itself:
\begin{eqnarray}
\widetilde{p} (\epsilon) &=& \frac{\epsilon  + w_r + \bar
w_d(\epsilon)}{\epsilon^2 + \epsilon [ w_r +2 \bar w_d(\epsilon)] +
w_r \bar w_d(\epsilon)}\nonumber\\
&=& \frac{1}{\epsilon + \bar w_d(\epsilon)(\epsilon + w_r)/[\epsilon
+ w_r + \bar w_d(\epsilon)]} \label{rel4},
\end{eqnarray}
where
\begin{equation} \label{rel5}
\bar w_d (\epsilon) = w_d + 2(\epsilon + w_{r\!p})v^2/[(\epsilon +
w_{r\!p})^2 + 4\varepsilon^2].
\end{equation}
and $w_{r\!p} = w_r + w_p$.

This expression enables us to analyze all possible effects of
relaxation in the case of Poissonian distribution. Formula
(\ref{rel4}) predicts different types of  $\widetilde{p} (\epsilon)$
dependence [and thus $p(t)$] on the average time between
measurements $\bar t = w_r^{-1}$. The important parameter, which
essentially determines the dependences of $\widetilde{p} (\epsilon)$
and $p(t)$ on $w_r$, is
\begin{equation} \label{rel6}
\bar w_d^0 = \bar w_d (0) = w_d + 2w_{r\!p}v^2/(w_{r\!p}^2 +
4\varepsilon^2).
\end{equation}
In particular, the parameter $t_Z$ is directly related to $\bar
w_d^0$:
\begin{equation} \label{rel6a}
t_Z = \widetilde{p} (0) = 1/w_r + 1/\bar w_d^0.
\end{equation}

Here we summarize some most interesting limiting $p(t)$ dependences
on $w_r$.

\paragraph{Slow population relaxation, $\bar w_d^{0} \ll w_r $.}

In the case of slow population relaxation (or fast repetition of
measurements as it is considered in Sec. IV C), when $\bar w_d^{0},
\ll w_r $ (implying that $v, w_d \ll w_r$), formula (\ref{rel4})
strongly simplifies predicting exponentially decreasing $p(t)$:
\begin{equation}\label{rel7}
\widetilde{p} (\epsilon) \approx (\epsilon + \bar
w_d^0)^{-1}\;\;\mbox{and} \;\; p(t) \approx e^{-\bar w_d^0 t},
\end{equation}
for which $t_Z = 1/\bar w_d^0$. Depending on the relation between
parameters of the system these expressions predict different
$w_r$-depedences of the RDM effect:

1) For fast dephasing, when $w_p \gg w_r$ and $w_{rp} \approx w_p$,
we find that the $p(t)$ decay rate and $t_Z$ are independent of
$w_r$, i.e. of the measurements, which can be considered as the
onset of the inverse Zeno regime \cite{Nam,Fac,Kos}: $\bar w_d^0
\approx w_d + 2v^2 w_p/(w_p^2 + 4\varepsilon^2) $.

2) For slow dephasing, when $w_p \ll w_r$, the decay rate $\bar
w_d^0 \approx w_d + 2v^2w_r/(w_r^2 + 4\varepsilon^2)$ and therefore
both quantum and inverse Zeno regimes are possible. For $w_r
> \varepsilon$ we get the decreasing function $\bar w_d^0 (w_r)
\approx w_d + 2v^2/w_r$, corresponding quantum Zeno case, while in
the opposite limit $w_r < \varepsilon$ we obtain the dependence
$\bar w_d^0 (w_r) \approx w_d + (v^2w_r)/(2\varepsilon^2)$, which is
associated with the inverse Zeno effect. Note that the above
analysis of $\bar w_d^0 (w_r)$ dependence (and RDM effect, in
general) is closely related to that of the rate $\bar w_0 (w_r)$ of
the (exponential) $p(t)$ decay in the absence of relaxation and for
fast repeated measurements ($w_r \gg v$), which is described by eq.
(\ref{poi3a}).

\paragraph{Fast population and phase relaxation $w_d, w_p \gg w_r $.}

In the limit of large population and phase relaxation rates $w_d,
w_p \gg w_r $ (recall that $w_p \geq w_d/2$) of special interest is
the case $\bar w_d^0 > w_r$, in which the evolution kinetics
consists of two stages: the stage of fast equilibration (at $t \sim
1/\bar w_d^0 $) and the stage of slow quasiequlibrium evolution
affected by measurement (at $t \geq 1/w_r > 1/\bar w_d^0 $). During
the first fast stage the survival probability $p (t)$ decreases from
$1$ to $1/2$ according to $p(t) \approx \frac{1}{2}(1 + e^{-2\bar
w_d^0 t})$. After that (during the second most interesting stage) $p
(t)$ decreases exponentially:
\begin{equation}\label{rel8}
\widetilde{p} (\epsilon) \approx \mbox{$\frac{1}{2}$}(\epsilon +
\bar w_r/2)^{-1}\;\;\mbox{and} \;\; p(t) \approx
\mbox{$\frac{1}{2}$} e^{- (w_r/2) t},
\end{equation}
so that  $t_Z = 1/w_r$ in agreement with the prediction of eq.
(\ref{rel6a}). Formulas (\ref{rel8}) can also be derived with the
use of eq. (\ref{sle6b4}) in which ${\widetilde{P}}_{p_0}
(\epsilon)$ and ${\widetilde{W}}_{p_0} (\epsilon)$ should be
replaced by ${\widetilde{P}}_{p_1} (\epsilon)$ and
${\widetilde{W}}_{p_1} (\epsilon)$, respectively, as it was
mentioned in the analysis of eq. (\ref{sle11d}). Taking into
consideration that in the limit of fast relaxation after short time
$\sim w_d^0$ the survival probability $p_1 (t) = \langle 1 |\rho
(t)| 1\rangle$ in the absence of measurements is given by $p_1 (t)
\approx 1/2$, we get ${\widetilde{P}}_{p_1}(\epsilon) =
\int_0^{\infty}\!dt \, e^{-\epsilon t} P(t)p_1 (t) \approx (\epsilon
+ w_r)^{-1}$ and ${\widetilde{W}}_{p_1}(\epsilon) =
\frac{1}{2}\int_0^{\infty}\!dt \, e^{-\epsilon t} W(t) p_1 (t)
\approx \frac{1}{2}w_r(\epsilon + w_r)^{-1}$. Substitution of these
expressions into eq. (\ref{sle6b4}) leads to formula (\ref{rel8}).

These formulas demonstrate that for $w_d, w_p \gg w_r $ the effect
of measurement shows itself in the inverse Zeno effect, in which the
effect increases as $w_r$ is increased.

\subsubsection{Anomalous distribution of measurements}

As in the case of dynamic systems, in our discussion of the
anomalous RDM effect in the presence of relaxation we will also
restrict ourselves to the analysis of the limit of short
characteristic time between measurements $w_r^{-1} \to 0$,
corresponding to $\|\hat \Omega_{\epsilon} \|/w_r \ll 1$.

Most clearly the effect of relaxation can be demonstrated in the
limit of fast dephasing, when $w_p \gg v$. In this limit the general
kinetic equation (\ref{rel1}) reduces to the system of balance
equations, i.e. equations for populations of states.

For the considered two level system (\ref{sle7}) with relaxation
superoperator (\ref{rel3}) the balance equations for the vector
${\bf p}(t)$ of populations of states $|1\rangle$ and $|2\rangle$,
can be written as
\begin{equation}\label{rel9}
\dot {\bf p} = -\hat R_d {\bf p}, \;\;\mbox{where} \;\; {\bf p}(t) =
p_1 (t)|11\rangle + p_2 (t)|22\rangle,
\end{equation}
and
\begin{equation} \label{rel10}
\hat R_d = \tilde w_d^0 (| 11\rangle - | 22 \rangle) (\langle 11 | -
 \langle 22 |)
\end{equation}
with 
\begin{equation} \label{rel11}
\tilde w_d^0 = w_d + 2v^2 w_{p}/(w_{p}^2 + 4\varepsilon^2).
\end{equation}

The function under study $\widetilde{p} (\epsilon) \equiv
\widetilde{p}_{\infty} (\epsilon)$, which is the Laplace transform
of the probability $p_{\infty} (t) \equiv p_1(t)$ of survival in the
state $|1\rangle$, can be calculated with the use of expression
(\ref{anom1a}) but with supermatrix $\hat \Omega_{\epsilon}$
replaced by $\hat {\bar \Omega}_{\epsilon} $ defined in the reduced
space of diagonal elements of the density matrix
\begin{eqnarray} \label{rel11a}
\widetilde{p}_{\infty} (\epsilon) &\approx& \langle 11|\hat {\bar
\Omega}_{d}^{\alpha -1}|11\rangle/\langle 11| \hat {\bar
\Omega}_{d}^{\alpha} |11\rangle\\
&=& \frac{\epsilon^{\alpha-1}+ (\epsilon + 2\tilde
w_d^0)^{\alpha-1}} {\epsilon^{\alpha}+ (\epsilon + 2\tilde
w_d^0)^{\alpha}},
\end{eqnarray}
where $\hat {\bar \Omega}_{d} = \epsilon + \hat R_d$.

This formula predicts the long tailed behavior $p_{\infty} (t) \sim
1/t^{\alpha}$ similar to that found in Sec. IVD for dynamic systems.
In the limit $\alpha \to 1$, however, the amplitude of the tail
becomes negligibly small and the kinetics reduces to the evident
exponential: $p_{\infty} (t) = e^{-\tilde w_d^0 t}$.

Similar to the dynamic systems, the most important feature of the
anomalous kinetics $p_{\infty} (t)$ in the presence of relaxation
consists in the independence of this kinetics of $w_r$, i.e. of the
characteristic time $w_r^{-1}$ between measurements. It is worth
noting, though, that this independence results from the existence of
the nontrivial special limit $w_r \to \infty$, manifesting itself in
eq. (\ref{anom1a}), rather than from the effect of relaxation
itself.

\section{Discussion and conclusions}

In this work we have discussed the RDM effect on the evolution of
quantum systems. The sequence of measurements is described as a
renewal stochastic process \cite{Wei,Bou}, whose specific properties
are controlled by the PDF $W(t)$ of time intervals $t$ between
measurements. The specific features of the RDM effect, which can be
called the stochastic Zeno effect, are essentially determined by the
analytical behavior of $W(t)$ and properties of the quantum system
under study, for example the presence of relaxation or decay
\cite{Fac,Kos,Fac1} (see below).

The analysis carried out in our work shows that in the case of
rapidly decreasing PDF $W(t)$ the RDM effect on dynamic systems
(without relaxation) is similar to that of measurements
equidistantly distributed in time \cite{Nak,Nam,Fac,Kos}. This
effect is characterized by the average time $\bar t =
\int_0^{\infty}\!dt\,t W(t)$ between measurements or the average
rate $w_r \sim 1/\bar t$ of the repetition of measurements. To
demonstrate the specific features of the effect of the RDM with
rapidly decreasing $W(t)$ on dynamic systems, we have analyzed the
effect on the two level system within the Poissonian model for the
RDM. The model is shown to allow one to describe both quantum and
inverse Zeno regimes of the RDM effect by very simple analytical
expressions.

In particular, it is possible to easily treat the main manifestation
of the quantum Zeno effect consisting in the reduction of the decay
rate of the measured state with the decrease of the time between
measurements. It is also shown that in the limit of frequently
repeated measurements $w_r \gg v$ the RDM result in the exponential
decay of the survival probability: $p(t) \approx e^{-\bar w_0 t}$
[eq. (\ref{poi3a})], with the rate $\bar w_0 (w_r)$ which is either
decreasing or increasing  function of $w_r$ depending on the
relation between $w_r$ and the splitting of levels $\varepsilon$.
These two type of behavior correspond to above-mentioned quantum and
inverse Zeno regimes, respectively \cite{Fac,Kos,Fac1}.

Of special interest is the anomalous case of heavy tailed PDF $W(t)
\sim 1/(w_r t)^{1+\alpha}, \: (\alpha < 1),$ in which the RDM effect
appears to strongly differ from that for rapidly decreasing PDF.
Note that in this case the average time $\bar t$ does not exist, but
the PDF is still characterized by the specific time $t_r =
w_r^{-1}$. For anomalous PDFs the Zeno effect is not observed: in
the limit $t_r \to 0$ the survival probability $p(t) \equiv
p_{\infty} (t)$ turns out to be a function of time $t$. Moreover the
probability $p_{\infty} (t)$ is a very slowly decreasing function:
$p_{\infty} (t) \sim 1/t^{\alpha}$.

The relaxation in the quantum system can strongly show itself in the
effect of the RDM on the evolution of the system. In particular, in
the case of rapidly decreasing $W(t)$, for not very large rates
$w_r$ smaller than the characteristic relaxation rate $\bar w_d^0$
(see Sec. V B), the survival probability $p(t)$ is demonstrated to
exponentially decrease $p(t) \sim e^{-\bar w_d^0 t}$ [eq.
(\ref{rel7})] with the rate $\bar w_d^0(w_r)$ either decreasing or
increasing as $w_r$ is increased. The form of $\bar w_d^0(w_r)$
behavior depends on the value of $(w_r + w_p)/\varepsilon$, where
$w_p$ is the dephasing rate. Similar to the pure dynamic case (in
the absence of relaxation) discussed above these types of
dependences can be considered as a manifestation of quantum and
inverse Zeno effects \cite{Fac,Kos,Fac1}. The acceleration of $p(t)$
decay by measurements, i.e. the inverse Zeno effect, is also found
in the limit of very fast relaxation $w_d, w_p \gg w_r$ [eq.
(\ref{rel8})].

In the case of anomalous $W(t)$ the relaxation in the system does
not lead to any new specific features of the evolution kinetics of
the system under study. Similar to the dynamic systems, for systems
with relaxation in the limit of high characteristic rate $w_r \to
\infty$ the survival probability $p_{\infty} (t)$ is still a
function of time slowly decreasing at large times: $p_{\infty} (t)
\sim 1/t^{\alpha}$.

In the end of this section we would like to discuss some recent
papers \cite{Kni,Bel,Rus} concerning the theoretical analysis of
quantum measurements and, in particular, the quantum Zeno effect
within the approaches which has something in common with those
applied in our work. In above-mentioned papers the kinetics of
measurement induced jump-like relaxation transitions in quantum
systems under study are described both numerically by direct Monte
Carlo modeling of stochastic quantum relaxation transition processes
\cite{Kni} and analytically by solving some Markovian kinetic
equations \cite{Bel,Rus} for corresponding PDFs. As far as
analytical approaches are concerned of particular interest is the
analysis of Poissonian-like theoretical models of stochastic
measurement induced jumps between states made in ref. \cite{Bel}. It
is worth noting though that the authors of this work restricted
themselves only to the most general (and at some respects too
formal) analysis of specific features of Poissonian jump processes,
including the equivalent description of the process in terms of the
SLE or corresponding Ito-type stochastic differential equations.
Interesting extension of the Markovian kinetic description of the
problem within a quantum model of the measuring device is considered
in ref \cite{Rus}. The perturbative treatment of the system-device
interaction allowed the authors to derive relatively simple
Markovian (Bloch-type) kinetic equations which describe the
evolution of the measurement affected system. Results obtained in
this work are, to some extent, related to some results of our study,
concerning the Poissonian RDM (since, as we have mentioned above,
the Markovian processes are directed related to Poissonian ones),
but unfortunately the direct comparison is hardly possible because
of difference in model parameters used in both treatments.

The important contribution of our work, as compared to those
discussed above, consists in the proposed new approaches, which
enable us to significantly generalize the analysis of measurement
effect on evolution of quantum systems, allowing for the
consideration of the RDM effects for different types of stochastic
RDM processes (described within the RA) from conventional Poissonian
one to anomalous. In addition, even in the Poissonian approach,
which is related to some Markovian models applied in early works,
the analysis is essentially extended by considering more thoroughly
the dephasing effect of measurements and by using more general
relaxation model for investigating the manifestation of relaxation
in the system.

Concluding our short analysis of obtained results we would like to
discuss the possibilities of experimental observation of the effects
predicted in our work. Of course, the question of great importance
is whether it is possible to realize the RDM experimentally.
Answering this question it is worth emphasizing the following
points:

1. The RDM can be realized experimentally with the use of equipment
which can randomize the set of measurements. Among possible variants
of random processes the Poissonian one is, probably, the most simple
for realization. At first sight the corresponding efforts will be of
not very much use in reality, however, it is worth noting that thus
obtained results on RDM effect can be analyzed somewhat easier than
those of equidistantly distributed measurements. The simplicity
results from the possibility of the analysis with simple Markovian
kinetic equations for any number of experiments. In addition, as it
was mentioned in the above discussion, the Poissonian RDM can be
realized in the measurement processes themselves due to stochastic
nature of the measurement procedure in some particular experiments
\cite{Kni,Bel,Rus}.

2. Perhaps, the most important fact is that the RDM can be realized
by the process under study itself. The point is that in many cases
the measurement is made within fairly small volume which can be a
small part of the larger volume where the process occurs. For
example, suppose that the measured system is a small Brownian
particle with quantum internal degrees of freedom, undergoing
stochastic motion within large volume. In its motion the particles
crosses the small measurement volume. Any visit of this volume can
be assumed to result in the measurement. In such a scenario of
measurements their statistics is represented by that of visits of
the small measurement site.

The statistics of visits depends on the mechanism of motion of the
particle. For instance, in the case of the particle, confined in
relatively small volume (of type of a cage) the statistics is close
to Poissonian. For freely diffusing motion in the infinite space the
statistics is quite well described by the anomalous variant of the
RA with the heavy tailed PDF $W(t) \sim 1/t^{1+\alpha}$, in which
$\alpha \leq 1 $ is determined by the dimensionality $d_s$ of the
space \cite{Shu1,Shu2}.

3. As a possible modification of the procedure, described above in
p. 2, one can consider the "measurement" of the quantum state of a
Brownian particle by another large quencher particle whose effect on
the quantum subsystem of the migrating Brownian particle can be
treated as a measuring device. The statistics of measurements in
such systems is, actually, determined that of reencounters of the
Brownian particle with the measuring quencher. The properties of
this statistics is well described by the RA, as mentioned above,
with the specific features depending on the mechanism of migration.


\textbf{Acknowledgements.}\, The author is grateful to Dr. V. P.
Sakun for valuable discussions. The work was partially supported by
the Russian Foundation for Basic Research.

\newpage

\textbf{Figure captions.}


Fig. 1. The dependence of the survival probability $p(t) \equiv
p(\tau|\tau_r)$ on the inverse average time between measurements
$\tau_r = \bar w_r^{-1} = v/w_r$ for two dimensionless times $\tau =
tv$ [$v$ is defined in eq. (\ref{sle8a})]: $\tau = 5$ (a) and $\tau
= 10$ (b). The probability is calculated for the two level system
(\ref{sle8a}), (\ref{sle8b}), assuming the Poissonian statistics of
measurements. Calculation is made with the use of formula
(\ref{poi2}) for four values of $\bar \varepsilon = \varepsilon/v $:
$\bar \varepsilon = 2.5$ (triangles), $\bar \varepsilon = 1.0$
(circles), $\bar \varepsilon = 0.5$ (squares), $\bar \varepsilon =
0.25$ (stars).

\medskip

Fig. 2. The dependence of the survival probability
$p_{\infty}(\tau)$ on $\tau = tv$, calculated by the inverse Laplace
transformation of $\widetilde{p}_{\infty}(\epsilon)$ (\ref{anom1b}),
for anomalous RDM with different parameters $\bar \varepsilon =
\varepsilon/v $ and $\alpha$: (a) $ \bar \varepsilon = 0.94,\:\alpha
= 0.1$ (full line); $ \bar \varepsilon = 0.53,\:\alpha = 0.3$
(dashed line); and (b) $ \bar \varepsilon = 0.53, \:\alpha = 0.92$
(1),\; $\bar \varepsilon = 0.94, \:\alpha = 0.92$ (2); $\bar
\varepsilon = 0.53, \:\alpha = 0.97$ (3); $\bar \varepsilon = 0.94,
\:\alpha = 0.97$ (4). In Fig. 2a the straight lines corresponding to
the dependence $p_{\infty}(t) = 1/(2.3 t^{\alpha})$ are presented
for the sake of demonstration of the asymptotic behavior of the
exact dependence $p_{\infty}(t)$. In fig. 2b the straight (dashed)
lines correspond to the approximate dependence (\ref{anom4}).

\newpage

\begin{figure}
\includegraphics[height=20.0truecm,width=14.0truecm]{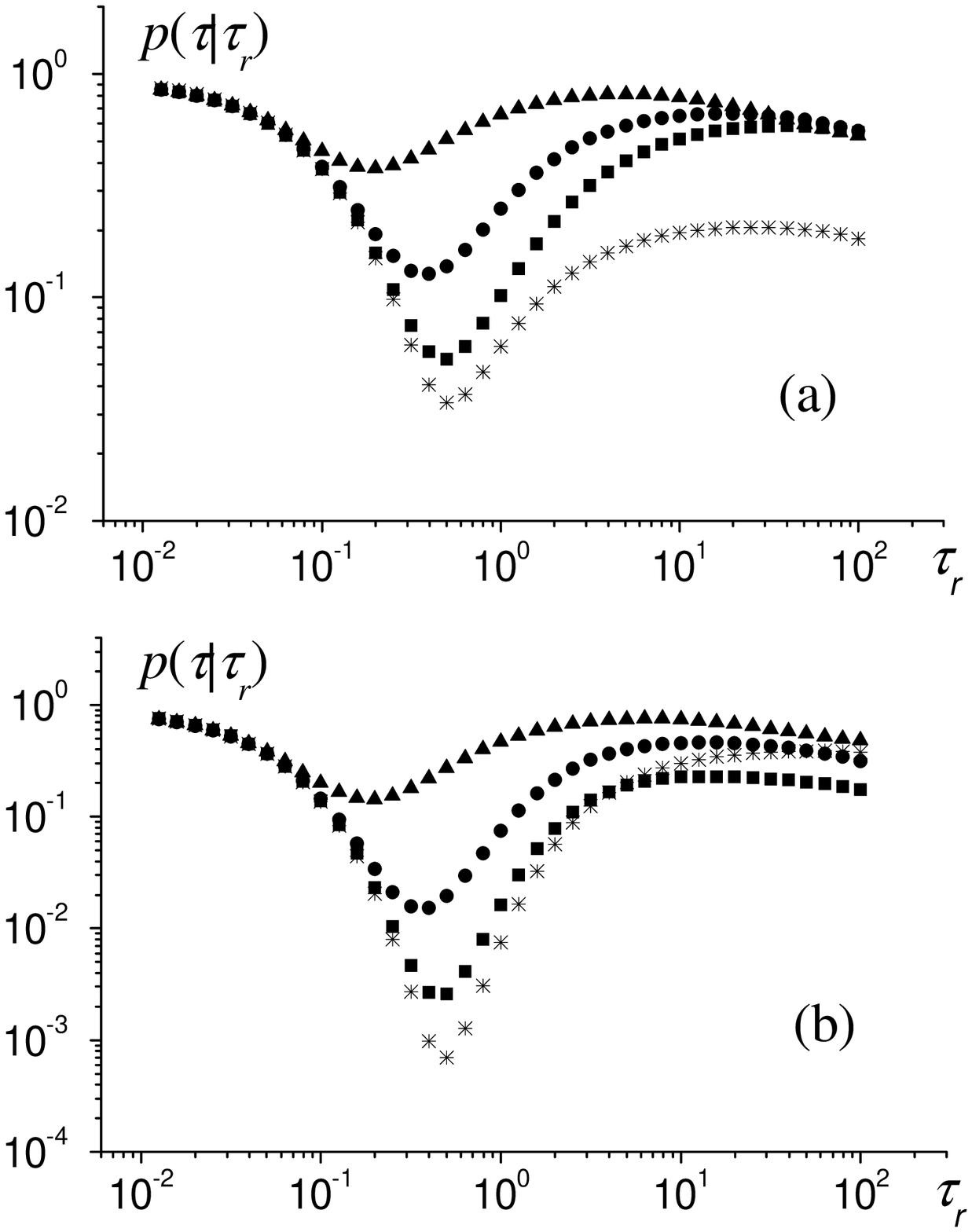}
\end{figure}

\center{{\textbf{{\Large Fig.1}}}

\newpage

\begin{figure}
\includegraphics[height=20.0truecm,width=14.0truecm]{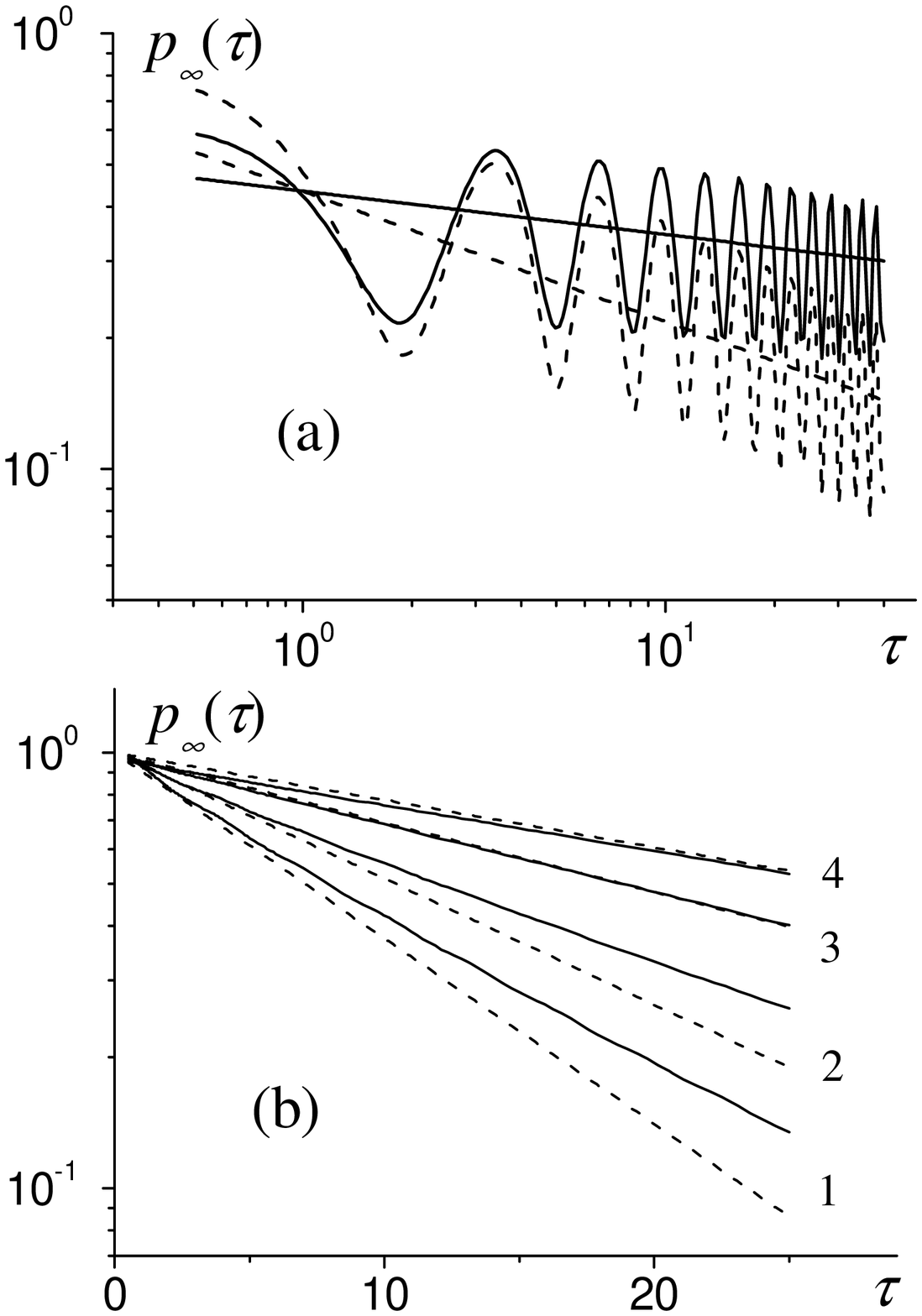}
\end{figure}

\center{{\textbf{{\Large Fig.2}}}

\end{document}